\documentclass[preprintnumbers,amsmath,amssymb]{revtex4}

\usepackage{mathrsfs}
\usepackage{amssymb}
\usepackage{graphicx}
\usepackage{dcolumn}
\usepackage{bm}
\usepackage{textcomp}
\usepackage{amsmath}
\usepackage[titletoc]{appendix}
\usepackage{accents}
\usepackage{ntheorem}
\usepackage{color}

\newcommand\ket[1]{\ensuremath{|#1\rangle}}
\newcommand\bra[1]{\ensuremath{\langle#1|}}
\newcommand\iprod[2]{\ensuremath{\langle#1|#2\rangle}}
\newcommand\oprod[2]{\ensuremath{|#1\rangle\langle#2|}}

\newcounter{RomanNumber}

\makeatletter
\def\widebar{\accentset{{\cc@style\underline{\mskip10mu}}}}
\def\Widebar{\accentset{{\cc@style\underline{\mskip8mu}}}}
\makeatother

\begin{document}
\title{Sending-or-not-sending twin-filed quantum key distribution with discrete phase modulation}

%
%
%
%

\author{Cong Jiang$ ^{1,2}$, Zong-Wen Yu$ ^{2,4}$, Xiao-Long Hu$^2$ and Xiang-Bin Wang$ ^{1,2,3,5\footnote{Email Address: xbwang@mail.tsinghua.edu.cn}\footnote{Also at Center for Atomic and Molecular Nanosciences, Tsinghua University, Beijing 100084, China}}$}

\affiliation{ 
\centerline{$^{1}$ Jinan Institute of Quantum technology, SAICT, Jinan 250101, China}
\centerline{$^{2}$State Key Laboratory of Low
Dimensional Quantum Physics, Department of Physics,} \centerline{Tsinghua University, Beijing 100084, China}
\centerline{$^{3}$ Synergetic Innovation Center of Quantum Information and Quantum Physics,}\centerline{University of Science and Technology of China, Hefei, Anhui 230026, China}
\centerline{$^{4}$Data Communication Science and Technology Research Institute, Beijing 100191, China}
\centerline{$^{5}$ Shenzhen Institute for Quantum Science and Engineering, and Physics Department,}
\centerline{Southern University of Science and Technology, Shenzhen 518055, China}}
\begin{abstract}
 We study the sending-or-not-sending (SNS) protocol with discrete phase modulation of coherent states. We first make the security of the SNS protocol with discrete phase modulation. We then present analytic formulas for key rate calculation. We take numerical simulations for the key rate through discrete phase modulation of both the original SNS protocol and the SNS protocol with two way classical communications of  active-odd-parity pairing (AOPP). Our numerical simulation results show that only with $6$ phase values, the key rates of the SNS protocol can exceed the linear bound, and with $12$ phase values, the key rates are very close to the results of the SNS protocol with continuously modulated phase-randomization.
\end{abstract}


\maketitle
\section{Introduction}
The theories and experiments of quantum key distribution (QKD)~\cite{bennett1984quantum,ekert1991quantum,pirandola1906advances,xu2019secure,gisin2002quantum,gisin2007quantum,
scarani2009security,shor2000simple,koashi2009simple,hwang2003quantum,
wang2005beating,lo2005decoy,wang2019practical,kraus2005lower,
yin2016measurement,liao2017satellite,boaron2018secure,chen2020sending,yin2020entanglement} have been widely studied since the first QKD protocol was proposed by Bennet and Brassard in 1984~\cite{bennett1984quantum}. The secure key rate and the distance are central issues in practical application of QKD. In particular, the decoy-state method~\cite{hwang2003quantum,wang2005beating,lo2005decoy} improves the relationship between key rate $R$ and channel transmittance $\eta$ from square scale, $R\sim O(\eta^2)$, to linear scale, $R\sim O(\eta)$. Recently, the idea of twin-field QKD (TFQKD)~\cite{lu2018overcoming} and its variants~\cite{wang2018twin,tamaki2018information,cui2019twin,curty2018simple,ma2018phase,lin2018simple,yu2019sending,maeda2019repeaterless,
lu2019twin,jiang2019unconditional,xu2019general,grasselli2019asymmetric,wang2019simple,hu2019general,zhang2019twin,zhou2019asymmetric} have further improved the key rate to the scale of square root of channel transmittance, $R\sim O(\sqrt{\eta})$, which can break linear bound~\cite{takeoka2014fundamental,pirandola2017fundamental} of QKD. So far, the TFQKD has been demonstrated by a number of experiments~\cite{minder2019experimental,liu2019experimental,wang2019beating,zhong2019proof,chen2020sending}.  

Among all the variants of TFQKD, the sending-or-not-sending (SNS) protocol~\cite{wang2018twin} together with its modified protocols~\cite{hu2019general,xu2019general,jiang2020zigzag} have attracted many attentions due to its large noise tolerance and high key rate.  Moreover, the SNS protocol has a unique advantage that the traditional decoy-state method directly applies, which makes the finite-key analysis very efficient. The SNS protocol has been experimentally demonstrated in proof-of-principle in Ref.~\cite{minder2019experimental}, and realized in real optical fiber with the finite-key effects taken into consideration~\cite{chen2020sending,liu2019experimental}. Notably, the SNS protocol has been experimentally demonstrated over 509 km optical fiber~\cite{chen2020sending} which is the longest secure distance of QKD in optical fiber.

In practice, we need the decoy-state method~\cite{hwang2003quantum,wang2005beating,lo2005decoy} to assure the security of those protocols with imperfect sources. In the traditional decoy-state method, phase randomization is requested so that the source state can be regarded as the classical mixture of different photon-number states. However, the perfect phase randomization by continuous modulation is technically not likely. In a real experiment, the phases of WCS sources are discretely modulated to $\frac{2m\pi}{N}$ where $m=0,1,2,\cdots,N-1$ and $N$ is always an even number. The major difference between the discrete modulation and continuous modulation is that the actual states of the latter case are Fock states while that of the former case is not. 

To close the gap between the theory and experiment, we study the SNS protocol with discrete phase modulation of WCS sources. Although the effects of discrete-phase-randomization have been studied for the protocols such as the traditional decoy-state BB84~\cite{hwang2003quantum,wang2005beating,lo2005decoy} in Ref.~\cite{cao2015discrete}, the MDIQKD~\cite{braunstein2012side,lo2012measurement,wang2013three,tamaki2012phase,
xu2013practical,
curty2014finite,xu2014protocol,
zhou2016making} in Ref.~\cite{cao2020discrete}, the non-post selection protocol~\cite{curty2018simple,cui2019twin} of TFQKD in Ref.~\cite{lorenzo2020twin}, no investigation has been done on the SNS protocol. Here we study this based on the structure of the SNS protocol. We prove it's security and present the formulas for key rate calculation. Unlike other protocols with discrete-phase-randomized WCS sources~\cite{cao2015discrete,cao2020discrete,lorenzo2020twin,zhang2020twin,primaatmaja2019versatile}. We then present analytical  formulas of the upper bound of the phase-flip error rate and the lower bound of the yield of untagged bits while the prior arte works have to solve linear programming problems. Our numerical simulation results show that only with 6 phase values, the key rates of the SNS protocol can exceed the PLOB bound, the linear bound of the key rate established by Pirandola, Laurenza, Ottaviani, and Banchi~\cite{pirandola2017fundamental}. With 12 phase slices, the key rates are very close to the SNS protocol with continuously modulated phase-randomized WCS sources. Since the property of no bit-flip error in the untagged bits in SNS protocl still holds with discrete phase modulation, we can directly apply the active-odd-parity pairing (AOPP) method proposed in our previous work~\cite{xu2019general} to improve the key rate. The numerical results show that the advantage of the AOPP method still holds.

The article is arranged as follows. We first introduce how to perform the SNS protocol with discrete phase modulation of WCS sources. Based on the equivalent entanglement protocol of the SNS protocol, we show how to get the formula of the phase-flip error rate. We then show how to apply the decoy-state method to obtain the upper bound of the phase-flip error rate and the lower bound of the yield of untagged bits. Using these  bounds, we present numerical results for both the original SNS protocol and the SNA protocol with AOPP method. The article is ended with the concluding remarks.

\section{The SNS protocol with discrete phase modulation of weak coherent state sources}
\subsection{The protocol}\label{protocol}
The implementation process of the SNS protocol with discrete phase modulation WCS sources is similar to that of the original SNS protocol~\cite{wang2018twin}. Here we first introduce the 4-intensity protocol as follows. Obviously, the special case that the intensity of signal state equals to that of one of the decoy state in the 4-intensity makes the 3-intensity protocol.

For each time window, Alice (Bob) randomly decides it is a decoy window or a signal window. If it is a decoy window, Alice (Bob) randomly chooses to prepare a pulse of state $\ket{0}$, $\ket{e^{2m\pi i/N}\sqrt{\mu_x}}$ or $\ket{e^{2m^\prime\pi i/N}\sqrt{\mu_y}}$
($\ket{0}$, $\ket{e^{2n\pi i/N}\sqrt{\mu_x}}$ or $\ket{e^{2n^\prime\pi i/N}\sqrt{\mu_y}}$), where $m,m^\prime,n,n^\prime$ are randomly chosen from $\{0,1,2,\cdots,N-1\}$ and $N$ is assumed to be an even number. If it is a signal window, Alice (Bob) randomly chooses to prepare a vacuum pulse or a pulse of state $\ket{e^{2l\pi i/N}\sqrt{\mu_z}}$ with probabilities $1-\epsilon$ and $\epsilon$ respectively, where $l$ is randomly chosen from $\{0,1,2,\cdots,N-1\}$. If Alice (Bob) decides to send out a vacuum pulse, that is to say sending nothing or not sending, Alice (Bob) takes the corresponding classical bit value as $0(1)$. If Alice (Bob) decides to send out a pulse of state $\ket{e^{2l\pi i/N}\sqrt{\mu_z}}$, Alice (Bob) takes the corresponding classical bit value as $1(0)$. 

Then Alice and Bob send their pulses to Charlie, Charlie is assumed to perform interferometric measurements on the received pulses. If only one of the two detectors clicks, Charlie would announce this pulse pair causes a click and whether the left detector or right detector clicks.  Alice and Bob take it as a one-detector heralded event.

After Alice and Bob repeat the above process for many times, they acquire a series of data which are used to perform the data post-processing. 

The first step of data post-processing is sifting. Alice and Bob first announce the types of each time window they have decided. For a window that both Alice and Bob have decided a signal window, it is a $Z$ window. The corresponding bits of the one-detector heralded event of the $Z$ windows, which are also called as the sifted key, are used to extract the final key. Except for the $Z$ windows, Alice and Bob announce the intensities and phases they have chosen in each window. For a window that both Alice and Bob have decided a decoy window, and the intensity of the pulse is $\mu_x$ and their phases satisfy
\begin{equation}
1-|\cos(\frac{2m\pi}{N}-\frac{2n\pi}{N})|\le \delta,
\end{equation}
where $\delta$ is a positive number close to 0, it is an $X_1$ window. Here in our calculation we shall simply set 
\begin{equation}
|\cos(\frac{2m\pi}{N}-\frac{2n\pi}{N})|=1
\end{equation} for the post selection condition of $X_1$ windows above. The data of $X_1$ windows are used to estimate the phase-flip error rate of untagged bits. The data of other windows are used for decoy-state analysis.

As the phases of WCS pulses in the $Z$ windows are never announced in the public channel, the density matrix of those WCS pulses is
\begin{equation}
\rho_z=\frac{1}{N}\sum_{l=0}^{N-1}\ket{e^{2l\pi i/N}\sqrt{\mu_z}}\bra{e^{-2l\pi i/N}\sqrt{\mu_z}}.
\end{equation}
For convenience, we define the approximated $j$-photon state $\ket{\lambda_j}$ in the following form~\cite{cao2015discrete}:
\begin{equation}\label{lambda1}
\ket{\lambda_j}=\frac{1}{\sqrt{P_j(\mu_z)}} \sum_{k=0}^{\infty}\frac{(\sqrt{\mu_z})^{kN+j}}{\sqrt{(kN+j)!}}\ket{kN+j},
\end{equation}  
and $j=0,1,2\cdots,N-1$. It is easy to see that $\rho_z$ is a classical mixture of different $\ket{\lambda_j}$. Explicitly, we have  
\begin{equation}
\rho_z=\sum_{j=0}^{N-1}P_j(\mu_z)\oprod{\lambda_j}{\lambda_j},
\end{equation}
where
\begin{equation}\label{pjz}
P_j(\mu_z)=\sum_{k=0}^{\infty}\frac{\mu_z^{kN+j}e^{-\mu_z}}{(kN+j)!}.
\end{equation}

For the pulse pairs in  $X_1$ windows, if the phases of the pulse pair satisfy $n\equiv (m+q) {\rm mod}N$ where $q$ is a constant integer, the density matrix of those pulse pairs is
\begin{equation}
\rho_{X1}(q)=\frac{1}{N} \sum_{m=0}^{N-1}\ket{e^{2m\pi i/N}\sqrt{\mu_x}}_A\ket{e^{2n\pi i/N}\sqrt{\mu_x}}_B\bra{e^{-2m\pi i/N}\sqrt{\mu_x}}_A\bra{e^{-2n\pi i/N}\sqrt{\mu_x}}_B,
\end{equation}
where the subscript $A$ and $B$ indicate Alice and Bob respectively. After a simple calculation, one can find that $\rho_{X1}(q)$ is actually the classical mixture of the state $\ket{\varphi_j^q}$, 
\begin{equation}
\rho_{X1}(q)=\sum_{j=0}^{N-1} PX_j(\mu_x)\ket{\varphi_j^q}\bra{\varphi_j^q}
\end{equation}
where 
\begin{equation}
\ket{\varphi_j^q}=\frac{e^{-\mu_x}}{\sqrt{PX_j(\mu_x)}} \sum_{k=0}^{\infty}\sum_{k_1=0}^{kN+j}\frac{(\sqrt{\mu_x})^{kN+j}}{\sqrt{k_1!(kN+j-k_1)!}}e^{\frac{2\pi i}{N}q(kN+j-k_1)}\ket{k_1;kN+j-k_1},
\end{equation}
and the probability to obtain the state $\ket{\varphi_j^q}$ is 
\begin{equation}
PX_j(\mu_x)=\sum_{k=0}^{\infty}\sum_{k_1=0}^{kN+j}\frac{{\mu_x}^{kN+j}e^{-2\mu_x}}{{k_1!(kN+j-k_1)!}} = \sum_{k=0}^{\infty} \frac{{(2\mu_x)}^{kN+j}e^{-2\mu_x}}{{(kN+j)!}} = P_j(2\mu_x).
\end{equation}

\subsection{The security analysis}
Follow the security proof in Ref.~\cite{wang2018twin}, let's first consider the equivalent entanglement protocol of the SNS protocol. 

In the entanglement protocol, for each time window, Alice and Bob pre-share the entanglement state
\begin{equation}
\begin{split}
\ket{\Psi_1}=&\frac{1}{\sqrt{2}}(\ket{0\lambda_1}\otimes \ket{\tilde{0}\tilde{0}}+\ket{\lambda_1 0}\otimes \ket{\tilde{1}\tilde{1}})\\
=&\frac{1}{\sqrt{2}}[\frac{1}{\sqrt{2}}(\ket{0\lambda_1}+\ket{\lambda_1 0})\otimes \frac{1}{\sqrt{2}}(\ket{\tilde{0}\tilde{0}}+\ket{\tilde{1}\tilde{1}})+\frac{1}{\sqrt{2}}(\ket{0\lambda_1}-\ket{\lambda_1 0})\otimes \frac{1}{\sqrt{2}}(\ket{\tilde{0}\tilde{0}}-\ket{\tilde{1}\tilde{1}})],
\end{split}
\end{equation} 
where $\ket{\tilde{0}\tilde{0}}$ and $\ket{\tilde{1}\tilde{1}}$ are local states that only exist in Alice's and Bob's labs and $\ket{0\lambda_1}$ and $\ket{\lambda_1 0}$ are real states that are sent to Charlie. According to the measurement results announced by Charlie, Alice and Bob can get a series of almost perfect entanglement state by applying entanglement purification to the local states. Two parameters are needed in the entanglement purification: the first is the bit-flip error rate in the $Z$ basis, $e_z$, and the second is the bit-flip error rate in the $X$ basis, $e^{ph}$, which is also the phase-flip error rate in the $Z$ basis. Here the $Z$ basis means $\{\ket{\tilde{0}},\ket{\tilde{1}}\}$, and the $X$ basis means $\{\frac{1}{\sqrt{2}}(\ket{\tilde{0}}+\ket{\tilde{1}}),\frac{1}{\sqrt{2}}(\ket{\tilde{0}}-\ket{\tilde{1}})\}$. Finally, by measuring the local states in the $Z$ basis, Alice and Bob can get secure final keys. 

As Alice and Bob only concern about the secure final keys, they needn't have to measure their local states after Charlie announces his measurement results, but they can just measure their local states before they send the real states to Charlie. If Alice and Bob each measures their local qubits in the $Z$ basis, it is equivalent to that Alice and Bob randomly send the pulse of state $\ket{0\lambda_1}$ or $\ket{\lambda_1 0}$ to Charlie. If Alice and Bob each measures their local qubits in the $X$ basis, it is equivalent to that Alice and Bob randomly send the pulse of state $\ket{\chi_0}=\frac{1}{\sqrt{2}}(\ket{0\lambda_1}+\ket{\lambda_1 0})$ or $\ket{\chi_1}=\frac{1}{\sqrt{2}}(\ket{0\lambda_1}-\ket{\lambda_1 0})$ to Charlie. As shown in Ref.~\cite{wang2018twin}, a phase error happens if Alice and Bob send $\ket{\chi_0}$ to Charlie and Charlie announces the right detector clicks or Alice and Bob send $\ket{\chi_1}$ to Charlie and Charlie announces the left detector clicks. 

Denote $T_0^R$ as the probability that Charlie announces the right detectors clicks while Alice and Bob send $\ket{\chi_0}$ to Charlie. Denote $T_1^L$ as the probability that Charlie announces the left detectors clicks while Alice and Bob send $\ket{\chi_1}$ to Charlie. Denote $s_1$ as the yield of $\frac{1}{2}(\ket{0\lambda_1}\bra{0\lambda_1}+\ket{\lambda_1 0}\bra{\lambda_1 0})$. We can calculate the phase-flip error rate by
\begin{equation}
e^{ph}=\frac{T_0^R+T_1^L}{2s_1}.
\end{equation}

According to the above discussion and the tagged-model, we can define the untagged bits in the real protocol as the bits in the $Z$ windows that Alice decides not to send and Bob actually sends a pulse of state $\oprod{\lambda_1}{\lambda_1}$ while Bob decides to send a WCS pulse with intensity $\mu_z$, or Bob decides not to send and Alice actually sends a pulse of state $\oprod{\lambda_1}{\lambda_1}$ while Alice decides to send a WCS pulse with intensity $\mu_z$. Finally, we can get the secure final key rate by
\begin{equation}
R=2\epsilon(1-\epsilon)P_1(\mu_z)s_1[1-H(e^{ph})]-S_zfH(E),
\end{equation} 
where $H(x)=-x\log_2(x)-(1-x)\log_2(1-x)$ is the Shannon entropy, $S_z$ is the yield of the events in the $Z$ windows, $f$ is error correction efficiency factor, and $E$ is the bit error rate of the sifted keys.

\subsection{The decoy-state method}
To clearly show how to apply the decoy-state method to this protocol, we denote Alice's sources $\ket{0}$, $\ket{e^{2m\pi i/N}\sqrt{\mu_x}}$, and $\ket{e^{2m^\prime\pi i/N}\sqrt{\mu_y}}$ by $o_A$, $x_A$, and $y_A$ respectively, and we also denote Bob's sources $\ket{0}$, $\ket{e^{2n\pi i/N}\sqrt{\mu_x}}$, and $\ket{e^{2n^\prime\pi i/N}\sqrt{\mu_y}}$ by $o_B$, $x_B$, and $y_B$. We denote the source of pulse pairs by $\kappa_A\varsigma_B$, where $\kappa,\varsigma=o,x,y$. And for simplicity, we omit the subscripts. For example, source $ox$ represents that Alice uses the source $o_A$ and Bob uses the source $x_B$. Without phase post-selection, the density matrices of sources $x_A,y_A,x_B,y_B$ have similar form with Eq.~\eqref{lambda1}. Specifically, we have
\begin{equation}
\rho_w=\sum_{j=0}^{N-1}P_j(\mu_w)\oprod{\lambda_j^w}{\lambda_j^w}, \quad (w=x,y),
\end{equation} 
where 
\begin{equation}
\ket{\lambda_j^w}=\frac{1}{\sqrt{P_j(\mu_w)}}\sum_{k=0}^{\infty}\frac{(\sqrt{\mu_w})^{kN+j}}{\sqrt{(kN+j)!}}\ket{kN+j},
\end{equation}
and $P_j(\mu_w)$ is defined in Eq.~\eqref{pjz}. Here $\rho_x$ is the density matrix of sources $x_A$ and $x_B$, and $\rho_y$ is the density matrix of sources $y_A$ and $y_B$. 

Denote the yield of sources $\kappa\varsigma$ by $S_{\kappa\varsigma}$. Denote the yield of states $\oprod{0\lambda_j^w}{0\lambda_j^w}$ and $\oprod{\lambda_j^w 0}{\lambda_j^w 0}$ by $Y_{vj}^w$ and $Y_{jv}^w$. In the original SNS protocol~\cite{wang2018twin}, as the continuously modulated phase-randomized WCS sources are used, we have
\begin{equation}
Y_{vj}^x=Y_{vj}^y,\quad Y_{jv}^x=Y_{jv}^y,
\end{equation}  
but this equality no longer holds in this protocol. Consider the properties of trace distance~\cite{cao2015discrete}, we have
\begin{equation}
|Y_{vj}^x-Y_{vj}^y|\le\sqrt{1-\left(F_{xy}^j\right)^2},\quad |Y_{jv}^x-Y_{jv}^y|\le\sqrt{1-\left(F_{xy}^j\right)^2},
\end{equation}
where
\begin{equation}
F_{xy}^j=\frac{|\iprod{\lambda_j^x}{\lambda_j^y}|}{\sqrt{\iprod{\lambda_j^x}{\lambda_j^x}\iprod{\lambda_j^y}{\lambda_j^y}}}=\frac{\sum_{k=0}^{\infty}\frac{(\mu_x\mu_y)^{(kN+j)/2}}{(kN+j)!}}{\sqrt{\sum_{k=0}^{\infty}\frac{\mu_x^{kN+j}}{(kN+j)!}\sum_{k=0}^{\infty}\frac{\mu_y^{kN+j}}{(kN+j)!}}},
\end{equation}
is the fidelity of states $\ket{\lambda_j^x}$ and $\ket{\lambda_j^y}$. 

Denote the yield of states $\ket{0\lambda_1}$ and $\ket{\lambda_10}$ by $s_{01}$ and $s_{10}$ respectively. And we have $s_1=\frac{1}{2}(s_{01}+s_{10})$. We can get the lower bound of $s_1$ by either analytical formula or the linear programming. 

Combining the equations
\begin{equation}
S_{ox}=\sum_{j=0}^{N-1}P_j(\mu_x)Y_{vj}^x=\sum_{j=0}^{N-1}P_j(\mu_x)Y_{vj}^y+\Delta, \quad S_{oy}=\sum_{j=0}^{N-1}P_j(\mu_y)Y_{vj}^y,
\end{equation} 
where
\begin{equation}
\Delta=\sum_{j=0}^{N-1}P_j(\mu_x)(Y_{vj}^x-Y_{vj}^y).
\end{equation}
we have
\begin{equation}
Y_{v1}^y=\frac{P_2(\mu_y)S_{ox}-P_2(\mu_x)S_{oy}-[P_0(\mu_x)P_2(\mu_y)-P_0(\mu_y)P_2(\mu_x)]Y_{v0}^y-P_2(\mu_y)\Delta-\xi}{P_1(\mu_x)P_2(\mu_y)-P_1(\mu_y)P_2(\mu_x)},
\end{equation}
where
\begin{equation}
\xi=\sum_{j=3}^{N-1}[P_j(\mu_x)P_2(\mu_y)-P_j(\mu_y)P_2(\mu_x)]Y_{vj}^y.
\end{equation}
It is easy to check that $\xi\le 0$ if the following condition holds
\begin{equation}\label{condition}
\frac{P_1(\mu_x)}{P_1(\mu_y)}\ge \frac{P_2(\mu_x)}{P_2(\mu_y)}\ge \frac{P_j(\mu_x)}{P_j(\mu_y)},\quad j=3,4,\cdots,N-1,
\end{equation}
which can be easily examined given values of $\mu_x$ and $\mu_y$. And in our numerical simulation, we found Eq.~\eqref{condition} always holds. With
\begin{equation}
Y_{v0}^y\le S_{oo}+\sqrt{1-F_{0}^2},\quad \Delta\le \Delta^U=\sum_{j=0}^{N-1}P_j(\mu_x)\sqrt{1-\left(F_{xy}^j\right)^2},\quad s_{01}\ge Y_{v1}^y-\sqrt{1-F_{1}^2},
\end{equation}
where
\begin{equation}
F_0=\frac{1}{\sqrt{\sum_{k=0}^{\infty}\frac{\mu_y^{kN}}{(kN)!}}},\quad F_1=\frac{\sum_{k=0}^{\infty}\frac{(\mu_y\mu_z)^{(kN+1)/2}}{(kN+1)!}}{\sqrt{\sum_{k=0}^{\infty}\frac{\mu_y^{kN+1}}{(kN+1)!}\sum_{k=0}^{\infty}\frac{\mu_y^{kN+1}}{(kN+1)!}}}.
\end{equation}
We have
\begin{equation}
s_{01}\ge s_{01}^L=\frac{P_2(\mu_y)S_{ox}-P_2(\mu_x)S_{oy}-[P_0(\mu_x)P_2(\mu_y)-P_0(\mu_y)P_2(\mu_x)]\left(S_{oo}+\sqrt{1-F_{0}^2}\right)-P_2(\mu_y)\Delta^U}{P_1(\mu_x)P_2(\mu_y)-P_1(\mu_y)P_2(\mu_x)}-\sqrt{1-F_{1}^2}.
\end{equation}

By similar method, we can prove that
\begin{equation}
s_{10}\ge s_{10}^L=\frac{P_2(\mu_y)S_{xo}-P_2(\mu_x)S_{yo}-[P_0(\mu_x)P_2(\mu_y)-P_0(\mu_y)P_2(\mu_x)]\left(S_{oo}+\sqrt{1-F_{0}^2}\right)-P_2(\mu_y)\Delta^U}{P_1(\mu_x)P_2(\mu_y)-P_1(\mu_y)P_2(\mu_x)}-\sqrt{1-F_{1}^2},
\end{equation}
if Eq.~\eqref{condition} holds. Finally, we have $s_1\ge \frac{1}{2}(s_{01}^L+s_{10}^L)$.

The remaining task is to estimate the upper bound of phase-flip error rate, $e^{ph}$, which is equivalent to estimate the upper bounds of $T_0^R$ and $T_1^L$. 

Denote $T_+^R$ as the probability that Charlie announces the right detector clicks while Alice and Bob send out the pulse pairs in the $X_1$ window and their phases satisfy $m=n$. Denote $T_-^L$ as the probability that Charlie announces the left detector clicks while Alice and Bob send out the pulse pairs in the $X_1$ window and their phases satisfy $m=(n+N/2){\rm mod}N$. Given the discussion above, we know that the pulse pairs in the $X_1$ window whose phases satisfy $m=n$ are the classical mixture of the state $\ket{\varphi_j^0}$. Thus by denoting $t_j^R$ as the probability that Charlie announces the right detector clicks while Alice and Bob send out the pulse pairs of state $\ket{\varphi_j^0}$, we have
\begin{equation}\label{t+r}
T_+^R=\sum_{j=0}^{N-1}PX_j(\mu_x)t_j^R.
\end{equation}
 
Denote $T_{00}$ and $T_{00}^\prime$ as the probability that Charlie announces the right detector clicks and left detector clicks while Alice and Bob send out a vacuum pulse pair. Consider the properties of trace distance, we have
\begin{equation}\label{t0r}
|t_0^R-T_{00}|\le \sqrt{1-F_{00}^2}, \quad |t_1^R-T_0^R|\le \sqrt{1-\left[F_{11}^{+}(0)\right]^2},
\end{equation}
where
\begin{equation}
F_{00}=\frac{|\iprod{\varphi_0^0}{0}|}{\sqrt{\iprod{\varphi_0^0}{\varphi_0^0}\iprod{0}{0}}}=\frac{1}{\sqrt{\sum_{k=0}^{\infty} \frac{{(2\mu_x)}^{kN}}{{(kN)!}}}},
\end{equation}
and
\begin{equation}\label{Im}
\begin{split}
&F_{11}^{+}(q)=\frac{|\iprod{\varphi_1^q}{\chi_0}|}{\sqrt{\iprod{\varphi_1^q}{\varphi_1^q}\iprod{\chi_0}{\chi_0}}}=\frac{\sqrt{Re_{+}^2+Im^2}}{\sqrt{\sum_{k=0}^{\infty}\frac{{(2\mu_x)}^{kN+1}}{{(kN+1)!}}\sum_{k=0}^\infty \frac{\mu_z^{kN+1}}{(kN+1)!}}},\\
&Re_+=\frac{1}{\sqrt{2}}\sum_{k=0}^\infty \frac{(\mu_x\mu_z)^{(kN+1)/2}}{(kN+1)!}[1+\cos\frac{2\pi}{N}q(kN+1)],\\
&Im=\frac{1}{\sqrt{2}}\sum_{k=0}^\infty \frac{(\mu_x\mu_z)^{(kN+1)/2}}{(kN+1)!} \sin\frac{2\pi}{N}q(kN+1).
\end{split}
\end{equation}

Combine Eqs.~\eqref{t+r} and \eqref{t0r}, we have
\begin{equation}
T_0^R\le T_0^{R,U}=\frac{T_+^R-PX_0(\mu_x)\left(T_{00}-\sqrt{1-F_{00}^2}\right)}{PX_1(\mu_x)}+\sqrt{1-\left[F_{11}^{+}(0)\right]^2}.
\end{equation}

With similar method, we have 
\begin{equation}
T_1^L\le T_1^{L,U}=\frac{T_-^L-PX_0(\mu_x)\left(T_{00}^\prime-\sqrt{1-F_{00}^2}\right)}{PX_1(\mu_x)}+\sqrt{1-\left[F_{11}^{-}(\frac{N}{2})\right]^2},
\end{equation}
where 
\begin{equation}
\begin{split}
&F_{11}^{-}(q)=\frac{|\iprod{\varphi_1^q}{\chi_1}|}{\sqrt{\iprod{\varphi_1^q}{\varphi_1^q}\iprod{\chi_1}{\chi_1}}}=\frac{\sqrt{Re_{-}^2+Im^2}}{\sqrt{\sum_{k=0}^{\infty}\frac{{(2\mu_x)}^{kN+1}}{{(kN+1)!}}\sum_{k=0}^\infty \frac{\mu_z^{kN+1}}{(kN+1)!}}},\\
&Re_-=\frac{1}{\sqrt{2}}\sum_{k=0}^\infty \frac{(\mu_x\mu_z)^{(kN+1)/2}}{(kN+1)!}[1-\cos\frac{2\pi}{N}q(kN+1)].
\end{split}
\end{equation}
and $Im$ has already been shown in Eq.~\eqref{Im}. Finally, we have
\begin{equation}
e^{ph}\le \frac{T_0^{R,U}+T_1^{L,U}}{s_{01}^L+s_{10}^L}.
\end{equation}
With all those formulas, we can now calculate the secure final key rate according to the observed values in the experiment.

We have obtained explicit formulas above for key rate calculation. In our calculation below, we shall use our analytical formulas above. Definitely, the key rate here can be also calculated through linear programming. It is
\begin{equation}
\begin{split}
\min\quad & s_1=\frac{1}{2}(s_{01}+s_{10}),\\
  s.t.\;\; \;\,\,&{\rm constraints\; with\; observed\; values\; of\;  number\; of\; post\; selected\; counts,\;e.g.}
\\
\quad & S_{ox}=\sum_{j=0}^{N-1}P_j(\mu_x)Y_{vj}^x,\quad S_{oy}=\sum_{j=0}^{N-1}P_j(\mu_y)Y_{vj}^y, \\
&S_{xo}=\sum_{j=0}^{N-1}P_j(\mu_x)Y_{jv}^x,\quad S_{yo}=\sum_{j=0}^{N-1}P_j(\mu_y)Y_{jv}^y,\\
& |S_{oo}-Y_{0v}^y|\le \sqrt{1-F_{0}^2},\quad |S_{oo}-Y_{v0}^y|\le \sqrt{1-F_{0}^2},\\
&|Y_{vj}^x-Y_{vj}^y|\le\sqrt{1-(F_{xy}^j)^2},\quad |Y_{jv}^x-Y_{jv}^y|\le\sqrt{1-(F_{xy}^j)^2},\\
& |s_{01}-Y_{v1}^y|\le \sqrt{1-F_{1}^2},\quad |s_{10}-Y_{1v}^y|\le \sqrt{1-F_{1}^2},
\end{split}
\end{equation} 
and so on.
After we get the lower bound of $s_1$ through linear programming, the remaining process is the same to the above method. 
\subsection {4-intensity protocol and 3-intensity protocol}
The 3-intensity protocol is simply a special case of our 4-intensity above by setting $\mu_z=\mu_y$.

\section{The numerical simulation}
In this part, we shall show some numerical simulation results.

We use the linear model to simulate the observed values~\cite{yu2019sending,jiang2019unconditional}. The distance between Alice and Charlie and the distance between Bob and Charlie is assumed to be the same. The properties of the two detectors of Charlie are assumed to be the same. The lowest intensities of the sources in the decoy window are set as $\mu_x\ge 0.001$ and $\mu_y\ge 0.002$, and the remaining parameters including $\mu_z$ and $\epsilon$ are optimized to obtained the highest key rates. The `Distance' shown in the figures of this part means the length of fiber between Alice and Bob. And the experiment parameters used in our numerical simulation are listed in Table.~\ref{exproperty}.

\begin{figure}[h]
\centering
\includegraphics[width=8cm]{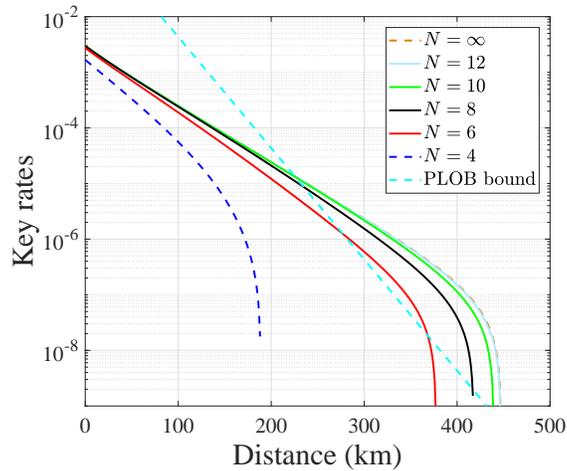}
\caption{The key rates of the SNS protocol with different number of phase values. The experiment parameters used in the numerical simulation are shown in Table.~\ref{exproperty}. The PLOB bound is used to measure the linear upper bound of the key rate of QKD~\cite{pirandola2017fundamental}.}\label{fig1}
\end{figure}

\begin{table}[h]
\begin{ruledtabular}
\begin{tabular}{ccccc}
$p_d$& $e_d$ &$\eta_d$ & $f$ & $\alpha_f$  \\
\hline
$1.0\times10^{-8}$& $3\%$  & $30.0\%$ & $1.1$ & $0.2$ \\ 
\end{tabular}
\end{ruledtabular}
\caption{List of experimental parameters used in numerical simulations. Here $p_d$ is the dark count rate of Charlie's detectors; $e_d$ is the misalignment-error probability; $\eta_d$ is the detection efficiency of Charlie's detectors; $f$ is the error correction inefficiency; $\alpha_f$ is the fiber loss coefficient ($dB/km$).}\label{exproperty}
\end{table}
Figure~\ref{fig1} is the key rate of the SNS protocol with different number of phase values. The cyan curve is the PLOB bound which is established by established by Pirandola, Laurenza, Ottaviani, and Banchi to measure the linear upper bound of the key rate of QKD~\cite{pirandola2017fundamental}. The key rates are almost coincide for the cases $N>12$. Thus the curves for $N>12$ are not listed in this figure. Figure~\ref{fig1} shows that only with $6$ phase values, the key rate of the SNS protocol can exceed the PLOB bound, and with $12$ phase values , the key rates is very close to the SNS protocol with continuously modulated phase-randomized WCS sources. 

\begin{figure}[h]
\centering
\includegraphics[width=8cm]{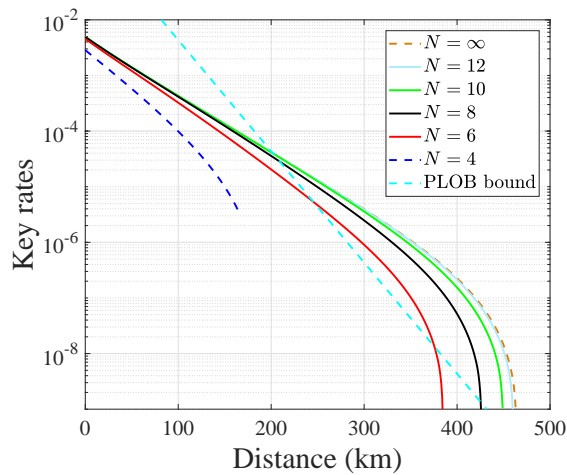}
\caption{The key rates of the SNS protocol with active-odd-parity pairing (AOPP)~\cite{xu2019general} and different number of phase values. The experiment parameters used in the numerical simulation are shown in Table.~\ref{exproperty}. The PLOB bound is used to measure the linear upper bound of the key rate of QKD~\cite{pirandola2017fundamental}.}\label{fig2}
\end{figure}

\begin{figure}[h]
\centering
\includegraphics[width=8cm]{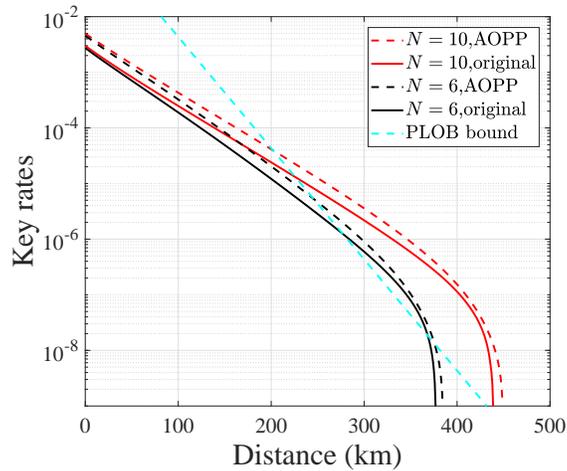}
\caption{The comparison of the key rates of the original SNS protocol and the active-odd-parity pairing (AOPP) method~\cite{xu2019general}. The experiment parameters used in the numerical simulation are shown in Table.~\ref{exproperty}. The PLOB bound is used to measure the linear upper bound of the key rate of QKD~\cite{pirandola2017fundamental}.}\label{fig3}
\end{figure}

Figure \ref{fig2} is the key rate of the SNS protocol with AOPP~\cite{xu2019general} and different number of phase values, while Figure \ref{fig3} is the comparison of the key rates of the original SNS protocol and the AOPP method. The AOPP method is an error rejection process through two way classical communication, which is only related to the sifted keys. As the property of no bit-flip error in the untagged bits holds in the SNS protocol with discrete phase modulaion of WCS sources, we can directly apply the formulas got in our previous work~\cite{xu2019general} to calculate the key rate after AOPP. Figure \ref{fig3} shows that key rates of the SNS protocol with AOPP exceed those of the original SNS protocol by about $70\%$ in all distances. While the distance between Alice and Bob is less than 150 km, the key rates of the AOPP method with 6 phase values are even higher than those of the original SNS protocol with 10 phase values.

\begin{figure}[h]
\centering
\includegraphics[width=8cm]{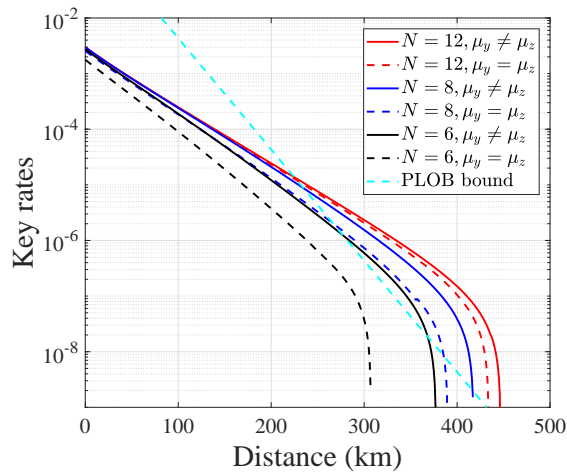}
\caption{The comparison of the key rates of the 4-intensity protocol ($\mu_y\neq \mu_z$) and the 3-intensity protocol ($\mu_y=\mu_z$). The experiment parameters used in the numerical simulation are shown in Table.~\ref{exproperty}. The PLOB bound is used to measure the linear upper bound of the key rate of QKD~\cite{pirandola2017fundamental}.}\label{fig4}
\end{figure}

Figure \ref{fig4} is the comparison of the key rates of the 4-intensity protocol and the 3-intensity protocol. In the 4-intensity protocol introduced in Sec.~\ref{protocol}, $\mu_y$ dosen't have to be equal to $\mu_z$. By adding a constraint that $\mu_y=\mu_z$, we get the 3-intensity protocol~\cite{yu2019sending}, which is a more convenient protocol in the experiments. Figure \ref{fig4} shows that while the number of phase values is large, which is closer to the case of continuously modulated protocol, the key rates of the 
3-intensity protocol is almost the same as those of the 4-intensity protocol. As the number of phase values decreases, the key rate gap between the two protocols gradually increases.

\section{Conclusion}
In summary, we have studied the SNS protocol with discrete phase modulation. Starting from the security proof, we obtain analytical formulas of the phase-flip error rate when the discrete phase modulation. With our derivations, we also get the lower bound of the yield of untagged bits.  Our numerical results show that only with $6$ phase values, the key rates of the SNS protocol can exceed the PLOB bound, and with $12$ phase values, the key rates are very close to the SNS protocol with continuously modulated phase-randomized WCS sources. The AOPP method proposed in Ref.~\cite{xu2019general} can be directly applied here, and the numerical results show that the advantage of the AOPP method still holds in the SNS protocol with discrete phase modulation of WCS sources.

\bibliography{refs}

\end{document}